\documentclass[10pt]{article}
\usepackage{amsmath,amssymb,amsfonts,amsthm,graphicx}
\title{Generalized coherent states for solvable quantum systems
 with degenerate discrete spectra and their nonclassical properties}
\author{G.R. Honarasa$^{1,2}$,
M.K. Tavassoly$^1$, M. Hatami$^1$ and R. Roknizadeh$^3$
\\
\footnotesize{$^1$ Atomic and Molecular Group, Faculty  of Physics, Yazd University, Yazd, Iran }
\\
\footnotesize{ $^2$ Physics Group, Faculty of Science, Shiraz University of Technology, Shiraz, Iran} \\
\footnotesize{ $^3$ Quantum Optics Group, Department of Physics, University of Isfahan, Isfahan, Iran}
\\ \footnotesize{e-mail: mktavassoly@yazduni.ac.ir  } }

\begin{document}

\maketitle \thispagestyle{empty}

 \begin{abstract}
  In this paper, the generalized coherent state for quantum systems with degenerate spectra is introduced. Then, the nonclassicality features and number-phase entropic uncertainty relation of two particular degenerate quantum systems are studied. Finally, using the Gazeau-Klauder coherent states approach, time evolution of some of the nonclassical properties of the coherent states corresponding to the considered physical systems are discussed.

 \end{abstract}

 {\bf keyword:}
   Solvable quantum system, Nonlinear coherent state, Phase property, Degenerate spectrum, Squeezing, Mandel parameter


\newcommand{\I}{\mathbb{I}}
\newcommand{\norm}[1]{\left\Vert#1\right\Vert}
\newcommand{\abs}[1]{\left\vert#1\right\vert}
\newcommand{\set}[1]{\left\{#1\right\}}
\newcommand{\R}{\mathbb R}
\newcommand{\C}{\mathbb C}
\newcommand{\DD}{\mathbb D}
\newcommand{\eps}{\varepsilon}
\newcommand{\To}{\longrightarrow}
\newcommand{\BX}{\mathbf{B}(X)}
\newcommand{\HH}{\mathfrak{H}}
\newcommand{\A}{\mathcal{A}}
\newcommand{\N}{\mathcal{N}}
\newcommand{\B}{\mathcal{B}}
\newcommand{\RR}{\mathcal{R}}
\newcommand{\HD}{\hat{\mathcal{H}}}

  \section{Introduction}\label{sec-intro}
 Generalized coherent states for solvable quantum systems with nondegenerate discrete energy spectra
 have been studied by us in \cite{honarasa}. The initial attempts to study the coherent states for
 quantum systems with degenerate discrete spectra were made by Kluader \cite{kluader}. Later, Crawford
 constructed generalized coherent states corresponding to these systems based on the Perelomov coherent
 states \cite{crawford}. Almost at the same time, Fox and Choi introduced another generalized coherent
 states for systems with degenerate energy spectra have been called Gaussian Klauder coherent
 states \cite{fox}. Then, Ali and Bagarello constructed vector coherent states of the Gazeau-Klauder type in the
 presence of degeneracies \cite{ali}. Coherent states for hydrogen atom as a degenerate system has been introduced in \cite{physica}. Recently, Dello Sbarba and Hussin studied
 new generalized and Guassian coherent states for quantum systems with degenerate spectra \cite{sbarba}. In the present paper, we will introduce generalized coherent states and study their nonclassicality features using the "nonlinear coherent states" approach when degeneracies in the energy spectra are present. \\
 The paper is organized as follows. In section 2, we introduce the general formalism for generalized coherent states corresponding to solvable quantum systems with degenerate spectra. The nonclassicality criteria of the obtained states are discussed in section 3 by Mandel parameter, quadrature evaluating squeezing and squeezing properties in phase or number operator, in addition to number-phase entropic uncertainty relation. Then, in Section 4 the presented approach is applied to a particle confined in a two-dimensional box and three-dimensional harmonic oscillator, and next the nonclassicality of their associated coherent states is investigated, numerically. Finally in section 5, with the help of Gazeau-Klauder coherent states formalism \cite{gazeau}, the dynamical evolution of the states will be introduced in section 2 are considered and based on this recognition \cite{roknizadeh,honarasa2}, the time evolution of nonclassical features and number-phase entropic uncertainty relation for the considered degenerate physical systems are studied.
 \section{Generalized coherent states for solvable quantum systems with degenerate spectra}
Recently, we have introduced the generalized coherent states for any arbitrary solvable quantum system in terms of its nondegenerate discrete spectrum, using the nonlinear coherent states approach as follows \cite{honarasa,roknizadeh}:
\begin{equation}\label{z,e}
   |z,e_n\rangle=\mathcal{N}_e(\left|z\right|^2)^{-1/2}\sum_{n=0}^\infty \frac{z^n}{\sqrt{[e_n]!}}|n\rangle,
\end{equation}
where
\begin{equation}\label{Ne}
   \mathcal{N}_e(\left|z\right|^2)=\sum_{n=0}^\infty \frac{\left|z\right|^{2n}}{[e_n]!},
\end{equation}
 and $[e_n]!=e_ne_{n-1}...e_1$ with $[e_0]!\doteq1$. Without loss of generality, it has been assumed that in the states (\ref{z,e}) the fundamental energy is zero and the others are in increasing order, i.e.,
\begin{equation}\label{spectra}
 e_0=0<e_1<e_2<...<e_{n-1}<e_n<...\;\;.
\end{equation}
But, clearly one can not use (\ref{z,e}) as the associated generalized coherent state for the Hamiltonian of physical system having a possibly degenerate energy spectrum like
\begin{equation}\label{degspectra}
 e_0\leq e_1\leq e_2\leq ...\leq e_{n-1}\leq e_n\leq ...\;\;\;.
\end{equation}
 To overcome the problem, following the Dello Sbarba and Hussin approach, by rearranging the degenerate energy spectrum (\ref{degspectra}) in increasing order one can redefine the corresponding spectrum as \cite{sbarba}
\begin{equation}\label{newspectra}
 \rho_0=0< \rho_1< \rho_2< ...<\rho_{n-1}< \rho_n< ...\;\;,
\end{equation}
 where each energy $\rho_n$ is degenerated $d_n$ times. It is worth to mention that the original spectrum has been shifted, so that the fundamental energy would be equal to zero. Now, based on the presented suggestion in \cite{ali,sbarba}, we pick an orthonormal basis $|n,p\rangle$ from the energy eigenstates ${|n\rangle}$, associated to each energy level $\rho_n$, where $p$ takes integer values between 0 and $d_n-1$ \cite{ali}. Also a real phase $\eta_n \in [0,2\pi]$ is introduced for each $n$. We thus associate with each energy $\rho_n$ a single state, constructed as a superposition of the states $|n,p\rangle$ as \cite{sbarba}
\begin{equation}\label{ndn}
   |n,d_n,\eta_n\rangle=\sum_{p=0}^{d_n-1} \exp{(ip\eta_n)}|n,p\rangle,
\end{equation}
 such that
\begin{equation}\label{nor}
 \langle n,d_n,\eta_n|m,d_m,\eta_m\rangle=d_n\delta_{nm}.
\end{equation}
 Note that the states $|n,d_n,\eta_n\rangle$ are orthogonal, but not normalized to unity. Now, we are ready to define the generalized coherent states for such quantum systems by
\begin{equation}\label{z,rho}
   |z,\{\rho_n\},\boldsymbol{\eta} \rangle=\mathcal{N}_{\rho}(|z|^2)^{-1/2}\sum_{n=0}^\infty \frac{z^n}{\sqrt{[\rho_n]!}}|n,d_n,\eta_n\rangle,
\end{equation}
with the normalization constant given by
\begin{equation}\label{Nrho}
   \mathcal{N}_{\rho}(|z|^2)=\sum_{n=0}^\infty \frac{\left|z\right|^{2n}}{[\rho_n]!}d_n,
\end{equation}
 where $[\rho_n]!=\rho_n\rho_{n-1}...\rho_1$ and the set $\{\rho_n\}$ contains $\rho_n,\; \rho_{n-1},\;...,\; \rho_1$. In the introduced state (\ref{z,rho}), the whole set of ${\eta_n}$ has been denoted by the infinite dimensional vector $\boldsymbol{\eta}$. It is worth to mention that in the presented approach, we supposed that the degeneracies are finite ($d_n<\infty$).
 \section{Nonclassical properties}
 In this section, some of the nonclassical properties of the introduced state will be discussed. For this purpose, we will discuss Mandel parameter, quadrature squeezing, number-phase squeezing of the general states (\ref{z,rho}). After all, the number-phase entropic uncertainty relation is also studied.
 \subsection{Mandel parameter}
 The number operator for physical systems with degenerate spectrum may be given by
\begin{equation}\label{number operator}
   N_D\equiv \sum_{n=1}^\infty \frac{n}{d_n} \left|n,d_n,\eta_n\rangle \langle n,d_n,\eta_n\right|.
\end{equation}
By the subscript $D$ (here and in the rest of the paper) we want to emphasize on the presence of the degeneracies. Clearly, the case $d_n=1$ recovers the usual number operator for nondegenerate case. It is easy to check that
\begin{equation}\label{nddn}
N_D|n,d_n,\eta_n\rangle=n|n,d_n,\eta_n\rangle.
\end{equation}
In this case the Mandel parameter may be defined as \cite{mandel}
\begin{equation}\label{Q}
Q_D=\frac{\langle N_D^2\rangle-\langle N_D\rangle^2}{\langle N_D\rangle}-1.
\end{equation}
With the help of (\ref{nddn}), one simply has
\begin{equation}\label{N}
\langle N_D\rangle\equiv \langle z,\{\rho_n\},\boldsymbol{\eta}|N_D|z,\{\rho_n\},\boldsymbol{\eta} \rangle=\mathcal{N}_{\rho}(|z|^2)^{-1}\sum_{n=0}^\infty  \frac{|z|^{2n}}{[\rho_n]!}nd_n,
\end{equation}
and similarly
\begin{equation}\label{N2}
\langle N_D^2\rangle\equiv \langle z,\{\rho_n\},\boldsymbol{\eta}|N_D^2|z,\{\rho_n\},\boldsymbol{\eta} \rangle=\mathcal{N}_{\rho}(|z|^2)^{-1}\sum_{n=0}^\infty  \frac{|z|^{2n}}{[\rho_n]!}n^2d_n.
\end{equation}
The $Q_D$ quantity determines the  quantum statistics of the radiation field states, i.e., it is super-Poissonian (if $Q_D > 0$), sub-Poissonian (if $Q_D < 0$) and Poissonian (if $Q_D = 0$).
 \subsection{Quadrature squeezing}
 We define the following annihilation and creation operators associate to any quantum systems with degenerate spectrum
 \begin{equation}\label{A}
 A_D=\sqrt{{H_\rho}_{n+1}}\sum_{n=0}^\infty \frac{1}{d_{n+1}}|n,d_n,\eta_n\rangle \langle n+1,d_{n+1},\eta_{n+1}|,
 \end{equation}
 \begin{equation}\label{A+}
 A_D^{\dagger}=\sqrt{{H_\rho}_n}\sum_{n=0}^\infty \frac{1}{d_{n+1}}|n+1,d_{n+1},\eta_{n+1}\rangle \langle n,d_n,\eta_n|,
 \end{equation}
 where two positive operators, ${H_\rho}_n$ and ${H_\rho}_{n+1}$ are distinctly defined as
\begin{equation}\label{HN}
 {H_\rho}_n\equiv \sum_{n=1}^\infty \frac{\rho_n}{d_n}|n,d_n,\eta_n\rangle \langle n,d_n,\eta_n|
 \end{equation}
 and
 \begin{equation}\label{HN+1}
 {H_\rho}_{n+1}\equiv \sum_{n=1}^\infty \frac{\rho_{n+1}}{d_n}|n,d_n,\eta_n\rangle \langle n,d_n,\eta_n|.
 \end{equation}
 The definitions (\ref{A}) and (\ref{A+}) may also be verified with the definition of Hermitian conjugate of operators. It is worth to notice that $(A_D^\dagger)^\dagger=A_D$. Indeed, we required this property which guarantees the hermiticity of the operators $\mathcal{X}$ and $\mathcal{P}$ which will be used in (\ref{xpu}). The following relations hold
 \begin{eqnarray}\label{An}
 A_D|n,d_n,\eta_n\rangle=\sqrt{\rho_n}|n-1,d_{n-1},\eta_{n-1}\rangle, \;\;\;\;\;\;\;\;\;\;\nonumber \\
 A_D^{\dagger}|n,d_n,\eta_n\rangle=\frac{d_n}{d_{n+1}}\sqrt{\rho_{n+1}}|n+1,d_{n+1},\eta_{n+1}\rangle,
 \end{eqnarray}
 and
 \begin{equation}\label{A+n}
 H_D|n,d_n,\eta_n\rangle=\rho_n|n,d_n,\eta_n\rangle,
 \end{equation}
 where $H_D$ is the Hamiltonian for degenerate system
 \begin{equation}\label{hd}
 H_D=\sum_{n=1}^\infty \rho_n \sum_{p=0}^{d_n-1}|n,p\rangle \langle n,p|.
 \end{equation}
 It is clear that the introduced states in (\ref{z,rho}) is the right eigenstate of annihilation operator defined in (\ref{An}), i.e., $A_D|z,\{\rho_n\},\boldsymbol{\eta} \rangle= z|z,\{\rho_n\},\boldsymbol{\eta} \rangle$. With the help of (\ref{An}) one can easily verify that the generalized coherent state (\ref{z,rho}) minimizes the Heisenberg uncertainty relation
  \begin{equation}\label{xpu}
 (\Delta \mathcal{X})^2(\Delta \mathcal{P})^2\geq\frac{1}{4}\left| \langle [\mathcal{X},\mathcal{P}] \rangle \right|^2,
 \end{equation}
 and we particularly have $(\Delta \mathcal{X})^2=(\Delta \mathcal{P})^2=\frac{1}{2}\left| \langle [\mathcal{X},\mathcal{P}] \rangle \right|$ with $\mathcal{X}=(A_D+A_D^\dagger)/\sqrt{2}$ and $\mathcal{P}=(A_D-A_D^\dagger)/i\sqrt{2}$ as two Hermitian operators. Therefore, no quadrature squeezing occurs for generalized coherent states associated to any degenerate quantum systems. Moreover, these states may be called as intelligent states \cite{tavassoly2,vogel}.
 \subsection{Phase properties and number-phase squeezing }
According to the Pegg-Barnett formalism, a complete set of $(s+1)$ orthonormal phase states $\theta_p$ may be defined by \cite{pegg}
\begin{equation}\label{tetaket}
   |\theta_m\rangle=\frac{1}{\sqrt{s+1}}\sum_{n=0}^s \exp{(in\theta_m)} |n\rangle,
\end{equation}
where $\{|n\rangle\}_{n=0}^s$ is the number states set with no degeneracy and $\theta_m$ is given by
 \begin{equation}\label{teta}
   \theta_m=\theta_0 +\frac{2\pi m}{s+1} ,\qquad   m=0,1,...,s,
 \end{equation}
 with arbitrary  value of $\theta_0$. The Hermitian phase operator based on the phase state (\ref{tetaket}), is then defined as
\begin{equation}\label{phi}
   \phi_\theta=\sum_{m=0}^s \theta_m |\theta_m\rangle \langle \theta_m|.
\end{equation}
Phase properties of nonlinear coherent states and coherent states associated to solvable quantum systems with nondegenerate spectra have been studied in the literature \cite{honarasa,physica2}.
Analogously to the phase state definition in (\ref{tetaket}), we can now introduce the following phase state and phase operator for a physical system with degenerate spectrum
\begin{equation}\label{thetaketnew}
   |\theta_m\rangle_D=\frac{1}{\sqrt{s+1}}\sum_{n=0}^s \frac{\exp{(in\theta_m)}}{\sqrt{d_n}} |n,d_n,\eta_n\rangle,
\end{equation}
\begin{equation}\label{phid}
   \phi_{\theta}^D=\sum_{m=0}^s {\theta_m |\theta_m\rangle_D}{_D\langle \theta_m|}.
\end{equation}
 Thus, the Pegg-Barnett phase probability distribution of the generalized coherent states may be defined as
\begin{equation}\label{p}
   P_D(\theta)=\lim_{s\rightarrow\infty}\frac{s+1}{2\pi} \left|{_D\langle\theta_m|z,\{\rho_n\},\boldsymbol{\eta}\rangle}\right|^2.
\end{equation}
 Using (\ref{z,rho}) and (\ref{thetaketnew}) in  (\ref{p}) one straightforwardly arrives at
 \begin{eqnarray}\label{pteta1}
    P_D(\theta)=\frac{\mathcal{N}_{\rho}(|z|^2)^{-1}}{2\pi}\left|\sum_{n} e^{-in\theta_n} \sqrt{d_n}\frac{z^n}{\sqrt{[\rho_n]!}}\right|^2 \;\;\;\;\;\;\;\;\;\;\;\;\;\;\;\;\;\;\;\;\;\;\;\;\;\;\;\;\;\;\;\;\;\;\;\;\;\;\;\;\;\;\;\;\;\;\;\;\;\;\; \\
  =\frac{\mathcal{N}_{\rho}(|z|^2)^{-1}}{2\pi}\left\{\sum_{n} \frac{|z|^{2n}}{[\rho_n]!}d_n  + 2\sum_{n} \sum_{k<n}\sqrt{\frac{d_n d_k}{[\rho_n]![\rho_k]!}} z^n z^{*k}   \cos{[(n-k)\theta]}\right\},\nonumber
\end{eqnarray}
 where the terms corresponding to $n=k$ and $n\neq k$ were written separately. Finally, with the help of (\ref{Nrho}) we will obtain
\begin{equation}\label{pthatanew}
   P_D(\theta)=\frac{1}{2\pi}\left(1+2\mathcal{N}_{\rho}(|z|^2)^{-1}\sum_{n} \sum_{k<n} \sqrt{\frac{d_nd_k}{[\rho_n]![\rho_k]!}}z^n z^{*k} \cos{[(n-k)\theta]}\right).
\end{equation}
 On the other side, the number and phase operators satisfy the following uncertainty relation
\begin{equation}\label{deltandeltaphi}
    \left\langle(\Delta\phi_{\theta}^D)^2\right\rangle \left\langle(\Delta N_D)^2\right\rangle \geq \frac {1}{4} |\langle [N_D,\phi_{\theta}^D] \rangle|^2,
\end{equation}
where analogously to nondegenerate case it may be easily proved that \cite{pegg}
\begin{equation}\label{nphi}
    [N_D,\phi_{\theta}^D]=i[1-2\pi P_D(\theta_0)].
\end{equation}
If $\left\langle(\Delta N_D)^2\right\rangle<\frac {1}{2} |\langle [N_D,\phi_{\theta}^D] \rangle|$ or $\left\langle(\Delta\phi_{\theta}^D)^2\right\rangle<\frac {1}{2} |\langle [N_D,\phi_{\theta}^D] \rangle|$ the squeezing occurs in number or phase operator, respectively. The phase variance for generalized coherent states (\ref{z,rho}) is given by
\begin{eqnarray}\label{deltaphi}
    \langle (\Delta\phi_{\theta}^D)^2\rangle=\int{\theta^2P_D(\theta)d\theta}-\left(\int{\theta P_D(\theta)d\theta} \right)^2 \;\;\;\;\;\;\;\;\;\;\;\;\;\;\;\;\;\;\;\;\;\;\;\;\;\;\;\;   \nonumber \\
    =\frac{\pi^2}{3}+4\mathcal{N}_{\rho}(|z|^2)^{-1}\sum_{n} \sum_{k<n} \sqrt{\frac{d_n d_k}{[\rho_n]![\rho_k]!}}z^n z^{*k} \frac{(-1)^{n-k}}{(n-k)^2},
\end{eqnarray}
in which (\ref{pthatanew}) is used. With the help of (\ref{N}) and (\ref{N2}), the number variance is given by
\begin{eqnarray}\label{deltan}
    \langle (\Delta N_D)^2\rangle=\langle N_D^2\rangle- \langle N_D\rangle ^2 \qquad \qquad \;\;\;\;\;\;\;\;\;\;\;\;\;\;\;\;\;\;\;\;\;\;\;\;\;\;\;\;\;\;\;\;\;\;\;\;\;\;\;\;\;\;\;\;\;\;\;\;\;\; \nonumber \\
    =\mathcal{N}_{\rho}(|z|^2)^{-1}\sum_{n} n^2  \frac{|z|^{2n}}{[\rho_n]!}d_n-\left(\mathcal{N}_{\rho}(|z|^2)^{-1}\sum_{n} n \frac{|z|^{2n}}{[\rho_n]!}d_n \right)^2.
\end{eqnarray}
To study the squeezing in number or phase components, the following number and phase squeezing parameters may be used
\begin{equation}\label{sn}
    S_N=\frac{2\langle (\Delta N_D)^2\rangle}{|\langle [N_D,\phi_{\theta}^D] \rangle|} -1,
\end{equation}
\begin{equation}\label{sphi}
    S_\phi=\frac{2\langle (\Delta\phi_{\theta}^D)^2\rangle}{|\langle [N_D,\phi_{\theta}^D] \rangle|} -1.
\end{equation}
If $S_N<0$ ($S_\phi<0$) then the associated state is number (phase) squeezed.
 \subsection{Number-phase entropic uncertainty relation }
 Let $\A$ and $\B$ be a pair of conjugate observables defined on an $(s+1)$-dimensional space, each with complete set of eigenstates $|a_n\rangle$ and $|b_n\rangle$ respectively, satisfying the eigenvalue equations \cite{honarasa2,joshi}
 \begin{equation}\label{eigenvalue2}
   \A|a_n\rangle=a_n|a_n\rangle,\;\;\;\; \B|b_n\rangle=b_n|b_n\rangle,
 \end{equation}
  where the discrete eigenvalues read as
 \begin{equation}\label{values}
   a_n=a_0+\frac{2\pi n}{(s+1)\beta},\;\;\;\; b_n=b_0+n\beta,
 \end{equation}
 with $a_0$, $b_0$ and $\beta$ as real constants.
 If the probability that a measurement of $\A$ gives the result $a_n$ is denoted by $P_a(a_n)$, with similar expression for $\B$, then the {\it "Shannon entropies"} associated with the probability distributions for $\A$ and $\B$ were defined as \cite{shannon}
 \begin{equation}\label{sa}
   S_{\A}=-\sum_{n=0}^s P_a(a_n)\ln P_a(a_n),
 \end{equation}
 \begin{equation}\label{sb}
   S_{\B}=-\sum_{n=0}^s P_b(b_n)\ln P_b(b_n).
 \end{equation}
 Kraus \cite{kraus} and Maassen \cite{maassen} found a uncertainty relation for these entropies which is dependent on the dimension of the space. Vaccaro et al. modified the definition of entropy and obtained a bound that is independent of dimension $s$ \cite{rojas}. Based on their proposal, the differences $\delta_a=2\pi/[(s+1)\beta]$ and $\delta_b=\beta$ respectively between successive eigenvalues of $\A$ and $\B$ are constants and so, they defined the new form of probability densities as
 \begin{equation}\label{papb}
   P_{\A}(n)=P_a(a_n)/\delta_a,\;\;\;\; P_{\B}(n)=P_b(b_n)/\delta_b.
 \end{equation}
 From the completeness and orthonormality of the eigenstates of $\A$ and $\B$ it follows that $\sum_{n=0}^s P_{\A}(n)\delta_a=1=\sum_{n=0}^s P_{\B}(n)\delta_b$. Hence, it is now possible to define new quantities $R_{\A}$ and $R_{\B}$,
 \begin{equation}\label{ra}
   R_{\A}\equiv -\sum_{n=0}^s \delta_a P_{\A}(n)\ln P_{\A}(n)=S_{\A}+\ln \delta_a,
 \end{equation}
 \begin{equation}\label{rb}
   R_{\B}\equiv -\sum_{n=0}^s \delta_b P_{\B}(n)\ln P_{\B}(n)=S_{\B}+\ln \delta_b.
 \end{equation}
 In the light of the above discussion Vaccaro et al. showed that working with $R_{\A}$ and $R_{\B}$ instead of $S_{\A}$ and $S_{\B}$ leads to following entropic uncertainty relation \cite{rojas}
 \begin{equation}\label{rarb}
   R_{\A}+R_{\B}\geq \ln(2\pi),
 \end{equation}
 which is independent of the dimension of the space. The equality holds for eigenstates of $\A$ and $\B$. \\
 By considering equations (\ref{number operator}), (\ref{teta}) and (\ref{phid}), we can identify general conjugate operators $\A$ and $\B$ with the phase and number operators, $\phi_{\theta}^D$ and $N_D$,  by choosing $\beta=1$, $a_0=\theta_0$ and $b_0=0$ in (\ref{values}). Henceforth, the entropic uncertainty relation for number and phase read as
 \begin{equation}\label{rphirn}
   R_\phi+R_N\geq \ln(2\pi).
 \end{equation}
 Now, we have all requirements to introduce the number-phase entropic uncertainty relations for generalized coherent states associated to solvable quantum systems with degenerate discrete spectra. In the infinite $s$ limit, the sum in (\ref{ra}) leads to following integral \cite{honarasa2,rojas}
 \begin{equation}\label{rphi}
   R_\phi=-\int_{\theta_0}^{\theta_0+2\pi} P_D(\theta)\ln P_D(\theta) d\theta,
 \end{equation}
 where $P_D(\theta)$ is the phase probability distribution obtained in (\ref{pthatanew}). On the other side the entropy of the number operator is defined as
 \begin{equation}\label{rn0}
   R_N=-\sum_{n=0}^\infty \left|\langle n,d_n,\eta_n|z,\{\rho_n\},\boldsymbol{\eta}\rangle \right|^2 \ln{(\left|\langle n,d_n,\eta_n|z,\{\rho_n\},\boldsymbol{\eta} \rangle \right|^2)}.
 \end{equation}
 By using (\ref{z,rho}) the right hand side of (\ref{rn0}) can be straightforwardly inverted to
 \begin{equation}\label{rn}
   R_N=-\mathcal{N}_{\rho}(\left|z\right|^2)^{-1} \sum_{n=0}^\infty \frac{|z|^{2n}}{[\rho_n]!}d_n\ln{\left(\mathcal{N}_{\rho}
   (\left|z\right|^2)^{-1}\frac{|z|^{2n}}{[\rho_n]!}d_n\right)},
 \end{equation}
 where $\mathcal{N}_{\rho}(\left|z\right|^2)$ has been defined in (\ref{Nrho}).

 \section{Some physical realizations of the formalism }
 Now, we are ready to apply the presented formalism to any physical system with known degenerate discrete spectra. We followed our research by applying the approach to two particular quantum systems: (i) a particle in a two-dimensional quantum square box and (ii) the three-dimensional harmonic oscillator.
        \subsection{A particle in a two-dimensional quantum square box}
The energy eigenvalues for a particle with mass $M$ in a two-dimensional quantum square box with sides $a$ are given by
\begin{equation}\label{enm}
e_{n,m}=\frac{\hbar^2\pi^2}{2Ma^2}(n^2+m^2)\equiv \hbar \omega_0(n^2+m^2),
\end{equation}
with $n,m\geq1$. We assume that $\hbar=1=\omega_0$ and so $e_{n,m}=n^2+m^2$. The first 23 different energy eigenvalues appearing in the spectrum are given in increasing order by:
\begin{eqnarray}\label{enm1}
\{2,\;5,\;8,\;10,\;13,\;17,\;18,\;20,\;25,\;26,\;29,\;32,\;34,\;37,\;40,\nonumber \\
41,\;45,\;50,\;52,\;53,\;58,\;61,\;65\}.\;\;\;\;\;\;\;\;\;\;\;\;\;\;\;\;\;\;\;\;\;\;\;\;\;\;\;\;\;\;\;\;\;\;\;\;
\end{eqnarray}
Rearranging and shifting the energy spectrum such that the fundamental energy being equal to zero we have indeed: $\rho_\nu=e_{n,m}-2=n^2+m^2-2$ which lead to the following spectrum:
\begin{eqnarray}\label{rhos}
\rho_0=0,\; \rho_1=3,\; \rho_2=6,\; \rho_3=8,\; \rho_4=11,\; \rho_5=15,\; \rho_6=16,\;\;\;\;\;\;\;\;\;\;\;\;\;\;\; \nonumber \\
\rho_7=18,\; \rho_8=23,\; \rho_9=24,\; \rho_{10}=27,\; \rho_{11}=30,\; \rho_{12}=32,\;\;\;\;\;\;\;\;\;\;\;\;\;\;\;\;\;\; \nonumber \\  \rho_{13}=35,\rho_{14}=38,\; \rho_{15}=39,\; \rho_{16}=43,\;\rho_{17}=48,\; \rho_{18}=50,\;\;\;\;\;\;\;\;\;\;\;\;\;\;\;\; \nonumber \\
\rho_{19}=51,\; \rho_{20}=56,\; \rho_{21}=59,\; \rho_{22}=63.\;\;\;\;\;\;\;\;\;\;\;\;\;\;\;\;\;\;\;\;\;\;\;\;\;\;\;\;\;\;\;\;\;\;\;\;\;\;\;\;\;\;\;\;\;\;
\end{eqnarray}
Recall that each energy $\rho_n$ has $d_n$-fold degeneracy. Corresponding to the set (\ref{rhos}), we thus have \cite{sbarba}
\begin{eqnarray}\label{dns}
d_0=1,\;d_1=2,\; d_2=1,\; d_3=2,\; d_4=2,\; d_5=2,\; d_6=1,\;\;\;\;\;\;\;\;\;  \nonumber \\
d_7=2,\; d_8=2,\; d_9=2,\; d_{10}=2,\; d_{11}=1,\; d_{12}=2,\;d_{13}=2,\;\;\;\;  \nonumber  \\
d_{14}=2,\; d_{15}=2,\; d_{16}=2,\; d_{17}=3,\; d_{18}=2,\;d_{19}=2,\; d_{20}=2, \nonumber  \\
d_{21}=2,\; d_{22}=4.\;\;\;\;\;\;\;\;\;\;\;\;\;\;\;\;\;\;\;\;\;\;\;\;\;\;\;\;\;\;\;\;\;\;\;\;\;\;\;\;\
;\;\;\;\;\;\;\;\;\;\;\;\;\;\;\;\;\;\;\;\;\;\;\;\;\;\;\;\;
\end{eqnarray}
Using the above set of numbers, we are interested in paying attention to the nonclassical properties of the associated coherent states may be constructed using (\ref{z,rho}). Now, we present our numerical result. Fig. 1 shows the Mandel parameter of the coherent states associated to a particle in a two-dimensional quantum box as a function
of $z \in \R$. The figure shows super-Poissonian statistics ($Q_D>0$) for $3.6\lesssim z\lesssim 5.7$ and sub-Poissonian statistics ($Q_D<0$) elsewhere. For $z\gtrsim 14$, Mandel parameter is almost equal to  $-1$, which is the minimum value corresponding to number states as the most nonclassical states. In Fig. 2 we have plotted the Pegg-Barnett phase distribution against $\theta$ for various fixed values of $z\in \R$  using (\ref{pthatanew}). From the figure it is seen that, $P_D(\theta)$ has a single peak at $\theta=0$, and as $z$ increases the peak becomes sharper and higher. To observe squeezing behavior in number or phase operators for the particular generalized coherent states we have plotted $S_N$ and $S_\phi$ against $z$ in Fig. 3. From the figure we find that, the squeezing parameters have opposite behavior, i.e., as $z$ increases $S_N$ increases while $S_\phi$ decreases. In detail, for $z \lesssim 1.2$, $S_N <0$ and $S_\phi>0$ implying squeezing occurrence in the number component. For $1.2\lesssim z\lesssim 1.9$ there is not any squeezing effect in neither of the conjugate components. But, as $z$ becomes greater than 1.9 we find that $S_\phi$ becomes negative indicating squeezing in the phase operator. In Fig. 4 we have plotted $R_{\phi}$, $R_N$ and their sum against $z \in \R$ for the associated coherent state. As it may be expected, the lower bound of the sum $R_N+R_\phi$ for all z according to (\ref{rphirn}) is satisfied. The sum $R_N+R_\phi$ is $\ln{(2\pi)}$ at $z = 0$ due to the fact that the vacuum is an eigenstate of number operator.
        \subsection{Three-dimensional harmonic oscillator}
The energy eigenvalues for three-dimensional harmonic oscillator in cartesian coordinates are given by
\begin{equation}\label{enmho}
e_{n,m,l}=(n+m+l+\frac{3}{2})\hbar \omega.
\end{equation}
According to the mentioned procedure we have the following expressions for the rearranged spectrum and number of degeneracies
\begin{equation}\label{rhooh}
\rho_{\nu}=n+m+l=\nu,
\end{equation}
\begin{equation}\label{dho}
d_\nu=\frac{(\nu+1)(\nu+2)}{2},
\end{equation}
which are necessary for constructing the coherent states for three-dimensional harmonic oscillator according to (\ref{z,rho}).
In writing (\ref{rhooh}) we have assumed $\hbar=1=\omega$.\\
Fig. 5 shows the Mandel parameter of the coherent states of the three-dimensional harmonic oscillator as a function of $z$. The sub-Poissonian statistics ($Q_D<0$) will be evident for all value of $z$. In Fig. 6 we have displayed the Pegg-Barnett phase distribution for the corresponding coherent states against $\theta$ for different fixed values of $z\in \R$. The behavior of $P_D(\theta)$ in this case is qualitatively similar to Fig. 2 for a particle in a two-dimensional box. From Fig. 7 we find that the squeezing parameters have opposite behaviors and show number or phase squeezing in small and large values of $z$, respectively. For the interval $0.4\leq z \leq 0.7$ squeezing effect neither in number nor in phase component is observed. Fig. 8 shows $R_{\phi}$, $R_N$ and their sum against $z \in \R$ for generalized coherent state associated to three-dimensional harmonic oscillator. It can be observed that the sum $R_N+R_\phi$ for all z is greater than $\ln{(2\pi)}$ and at $z = 0$ it gets it's minimum value due to the fact that the vacuum is an eigenstate of number operator.\\
We also applied the presented formalism to two-dimensional harmonic oscillator where
$\rho_{\nu}=n+m=\nu$ and $d_\nu=\nu+1$. The results are qualitatively the same as what
obtained for three-dimensional harmonic oscillator and therefore we will not bring them here.
    \section{Gazeau-Klauder coherent states associated to solvable quantum systems with degenerate discrete spectra}
 At this point we would like briefly noticing the link between Gazeau-Klauder coherent states \cite{gazeau} and the states introduced in (\ref{z,rho}). Upon the proposal introduced in \cite{roknizadeh}, by the action of the evolution type operator
 \begin{equation}\label{eval}
   S_D(\gamma)= e^{-i(\frac{\gamma}{\hbar \omega})H_D},
 \end{equation}
 on the states in (\ref{z,rho}), one can obtain the analytical form of Gazeau-Klauder coherent states for discrete and degenerate spectra as
 \begin{equation}\label{zgamma}
   |z,\gamma,\{\rho_n\},\boldsymbol{\eta} \rangle=\mathcal{N}_{\rho}(\left|z\right|^2)^{-1/2}\sum_{n=0}^\infty \frac{z^n e^{-i\rho_n\gamma}}{\sqrt{[\rho_n]!}}|n,d_n,\eta_n\rangle,
 \end{equation}
 where $\gamma\in \R$ and we have assumed $\hbar=1=\omega$. These set of states is the same as generalized coherent states which introduced by Dello Sbarba and Hussin \cite{sbarba}. Although the coherent states introduced by Ali and Bagarello in \cite{ali} seem to be similar to our introduced states, but they are not exactly the same. Indeed, the difference comes out from the fact that our basis $|n,d_n,\eta_n\rangle$ which have been used for the construction of the states (\ref{z,rho}) are not normalized (see equation (\ref{nor})). The introduced states in (\ref{zgamma}) are the right eigenstates of the following annihilation operator
\begin{equation}\label{AG}
 A_D^{GK}=\sqrt{{H_\rho}_{n+1}}e^{i[{H_\rho}_{n+1}-{H_\rho}_n]\gamma}\sum_{n=0}^\infty \frac{1}{d_{n+1}}|n,d_n,\eta_n\rangle \langle n+1,d_{n+1},\eta_{n+1}|,
 \end{equation}
 The corresponding creation operator is obtained as
 \begin{equation}\label{A+G}
 {A_D^{GK}}^{\dagger}=\sqrt{{H_\rho}_n}e^{-i[{H_\rho}_n-{H_\rho}_{n-1}]\gamma}\sum_{n=0}^\infty \frac{1}{d_{n+1}}|n+1,d_{n+1},\eta_{n+1}\rangle \langle n,d_n,\eta_n|,
 \end{equation}
 where ${H_\rho}_n$ and ${H_\rho}_{n+1}$ were introduced in (\ref{HN}) and (\ref{HN+1}).
Setting $\gamma\equiv t$ in (\ref{eval}) one can interpret the states in (\ref{zgamma}) as the time evolution of the states in (\ref{z,rho}) \cite{roknizadeh}. It can be easily observed that $\exp{(-iH_Dt)}|z,\gamma,\{\rho_n\},\boldsymbol{\eta} \rangle=|z,\gamma+t,\{\rho_n\},\boldsymbol{\eta} \rangle$, i.e., these states are temporally stable. The state (\ref{zgamma}) provides us with a powerful and at the same time simple formalism to investigate the nonclassical properties of state (\ref{z,rho}) as time goes on. Making use of (\ref{zgamma}) and following the same procedure leads to (\ref{N}), (\ref{N2}) and (\ref{rn}), exactly the same results for $Q_D$, $S_N$ and $R_N$ are again obtained. But, for the phase distribution of the temporally stable states (\ref{zgamma}), the relation (\ref{pthatanew}) straightforwardly changes to
 \begin{eqnarray}\label{ptetagk}
 P_D(\theta,\gamma)=\frac{1}{2\pi}(1+2\mathcal{N}_{\rho}(|z|^2)^{-1}\sum_{n} \sum_{k<n} \sqrt{\frac{d_nd_k}{[\rho_n]![\rho_k]!}}z^n z^{*k} \nonumber \\
 \times \cos{[(n-k)\theta]\cos{[(\rho_n-\rho_k)\gamma]}}).  \;\;\;\;\;\;\;\;\;\;\;\;\;\;\;\;\;
 \end{eqnarray}
At last, after using (\ref{ptetagk}) in (\ref{deltaphi}) and (\ref{rphi}) the following expressions for the phase variance and entropy associated with the phase operator can be obtained, respectively
 \begin{eqnarray}\label{deltaphigk}
    \langle (\Delta\phi_{\theta_D})^2\rangle=\int{\theta^2P_D(\theta,\gamma)d\theta}-\left(\int{\theta P_D(\theta,\gamma)d\theta} \right)^2 \;\;\;\;\;\;   \nonumber \\
    =\frac{\pi^2}{3}+4\mathcal{N}_{\rho}(|z|^2)^{-1}\sum_{n} \sum_{k<n} \sqrt{\frac{d_n d_k}{[\rho_n]![\rho_k]!}}z^n z^{*k}\nonumber \\
    \times \frac{(-1)^{n-k}}{(n-k)^2}\cos{[(\rho_n-\rho_k)\gamma]},\;\;\;\;\;\;\;\;\;\;\;\;\;\;\;\;\;\;\;\;\;\;\;
\end{eqnarray}
 \begin{equation}\label{rphigk}
   R_\phi(\gamma)=-\int_{\theta_0}^{\theta_0+2\pi} P_D(\theta,\gamma)\ln P_D(\theta,\gamma) d\theta.
 \end{equation}
 For the physical appearance of the latter states, we apply it to a particle in a two-dimensional quantum box and three-dimensional harmonic oscillator. In Fig. 9a we have plotted the sum $R_N+R_\phi$ against $z\in \R$ and $\gamma$ (dimensionless time) for a particle in a two-dimensional quantum box. Fig. 9b is the cross sections of Fig. 9a for some various fixed values of $z\in \R$. The figure shows oscillatory behavior of the quantity as $\gamma$ increases.
 We have plotted the sum $R_N+R_\phi$ against $z$ and $\gamma$ for the three-dimensional harmonic oscillator in Fig. 10a. It's oscillatory behavior with $\gamma$ is clearly seen. The cross section of Fig. 10a has been shown in Fig. 10b for some fixed values of $z\in \R$.

    \section{Summary and conclusion}
 We have introduced the generalized coherent states for solvable quantum systems with 
 degenerate discrete spectra. Then, we studied their nonclassicality features through 
 discussing the quantum statistical properties of the obtained states by paying attention to 
 Mandel parameter. Also, quadrature squeezing has been discussed. Although we found no quadrature 
 squeezing, but the states behave like the well-known intelligent states irrespective of 
 the characterizations of the quantum systems. Number and phase squeezing, as well as 
 number-phase entropic uncertainty relation of these states have been investigated through
  the Pegg-Barnett formalism. Next, the proposal has been applied to a few particular 
  physical systems with known degenerate spectra  and their nonclassical properties 
  have been investigated, numerically. It is found that, the generalized coherent states 
  for considered physical systems obey sub-Poissonian statistics for some intervals of
   $z\in \R$ and exhibit phase or number squeezing for different ranges of real values 
   of $z$. Also, they satisfy number-phase entropic uncertainty relation for all values 
   of $z$. At last, we study the time evolution of nonclassical properties of the particular
    states associated to physical systems, with the help of the temporally stable Gazeau-Klauder
     coherent states approach. We found that the sum of the entropies of the number and phase operators, 
     satisfy the entropic uncertainty relation and  
       oscillates with time.
 
 \newpage

\begin{figure}
	\centering
		\includegraphics[width=0.80\textwidth]{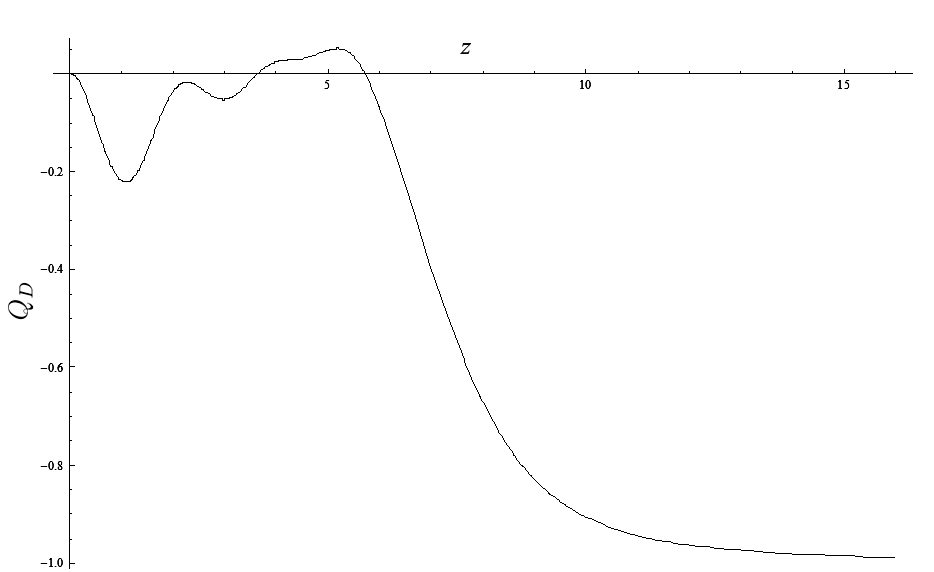}
	\caption{Plot of Mandel parameter of the generalized coherent states versus $z$ for a particle in a two-dimensional quantum square box.}
	\label{fig:figure1}
\end{figure}

\begin{figure}
	\centering
		\includegraphics[width=0.80\textwidth]{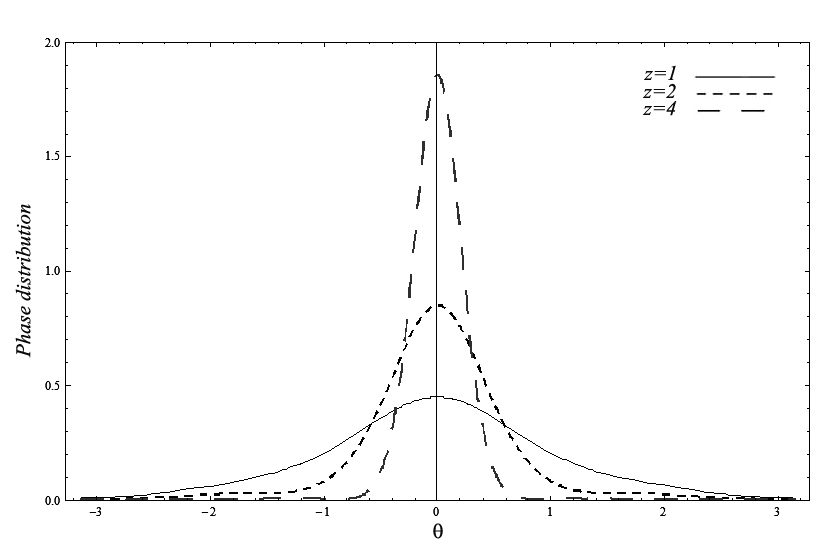}
	\caption{Pegg-Barnett phase distribution against $\theta$ for various fixed values of $z \in \R$ for a particle in a two-dimensional quantum square box.}
	\label{fig:figure2}
\end{figure}
\begin{figure}
	\centering
		\includegraphics[width=0.80\textwidth]{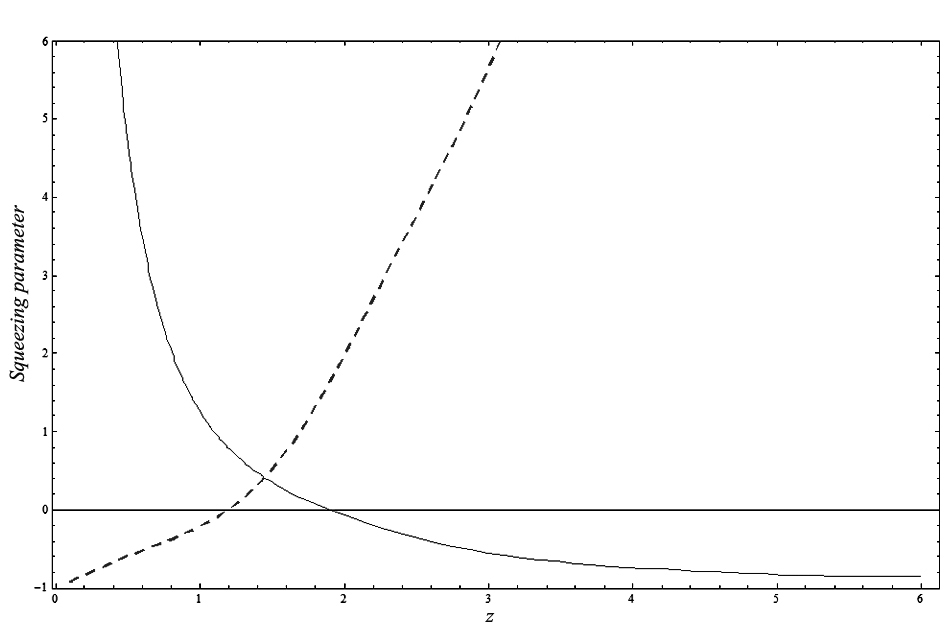}
	\caption{Plots of $S_\phi$ (solid curve), $S_N$ (dashed curve) against $z\in \R$ for a particle in a two-dimensional quantum square box.}
	\label{fig:figure3}
\end{figure}

\begin{figure}
	\centering
		\includegraphics[width=0.80\textwidth]{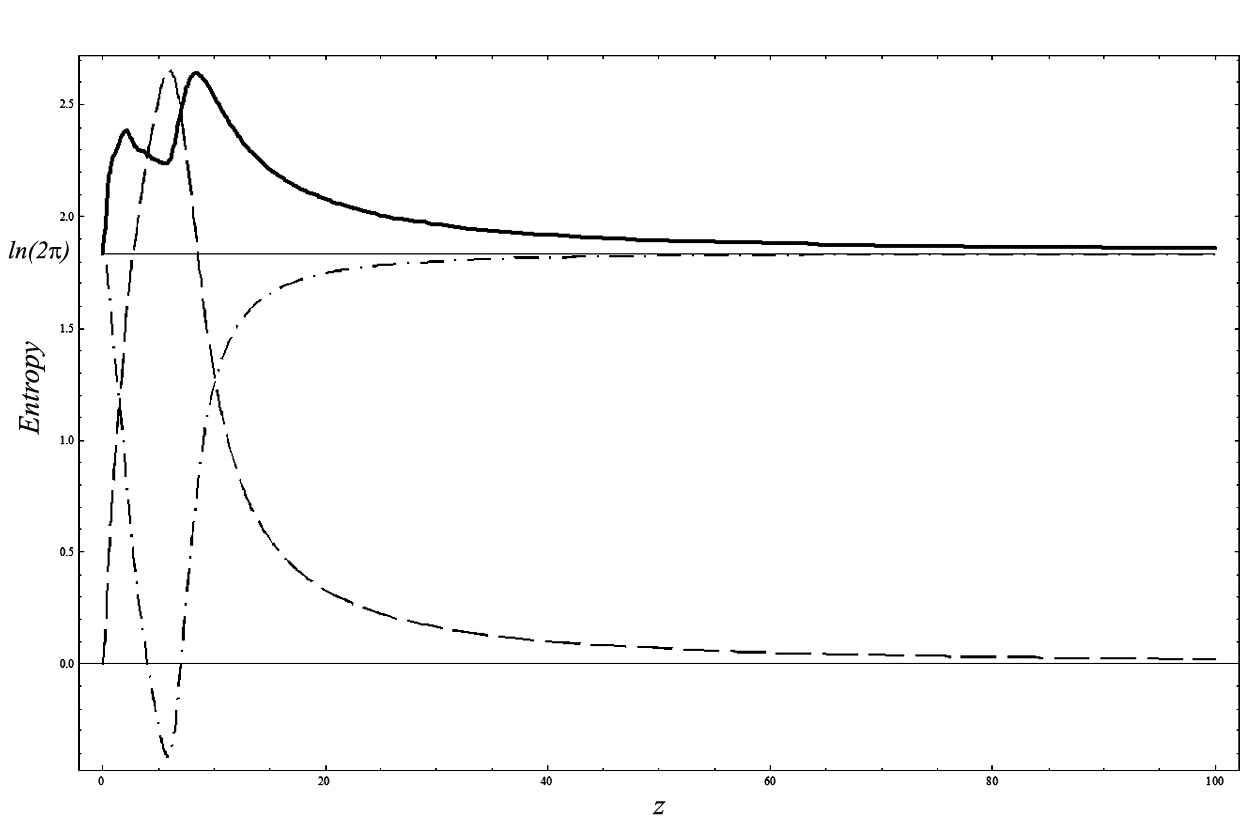}
	\caption{Plots of $R_\phi$ (dot-dashed curve), $R_N$ (dashed curve) and their sum (solid curve) against $z\in \R$ for a particle in a two-dimensional quantum square box.}
	\label{fig:figure4}
\end{figure}
\begin{figure}
	\centering
		\includegraphics[width=0.80\textwidth]{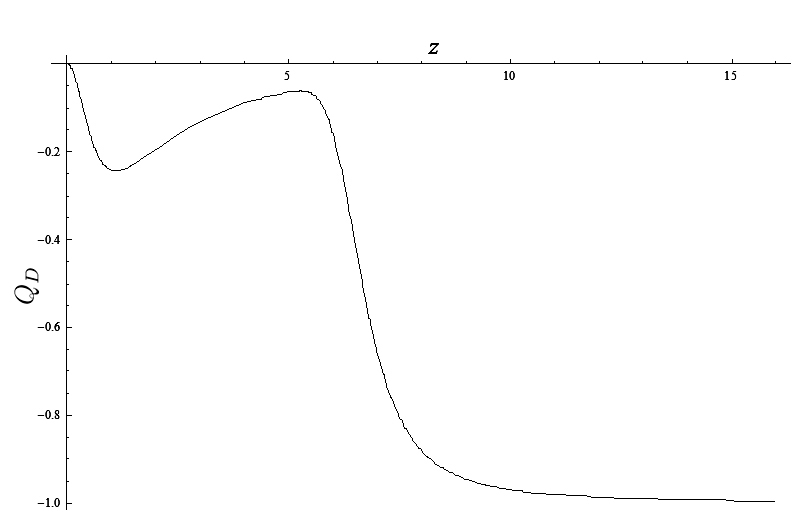}
	\caption{Plot of Mandel parameter of the generalized coherent states versus $z$ for three-dimensional harmonic oscillator.}
	\label{fig:figure5}
\end{figure}
\begin{figure}
	\centering
		\includegraphics[width=0.80\textwidth]{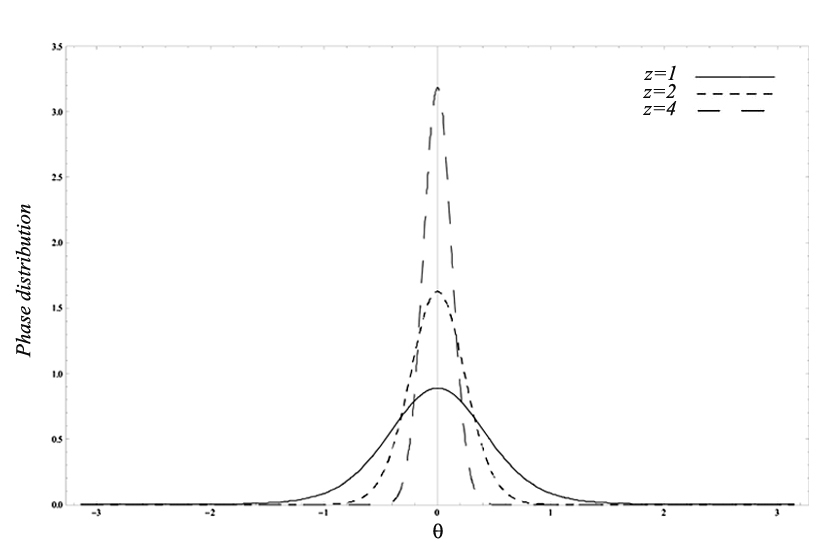}
	\caption{Pegg-Barnett phase distribution against $\theta$ for various fixed values of $z \in \R$ for three-dimensional harmonic oscillator.}
	\label{fig:figure6}
\end{figure}

\begin{figure}
	\centering
		\includegraphics[width=0.80\textwidth]{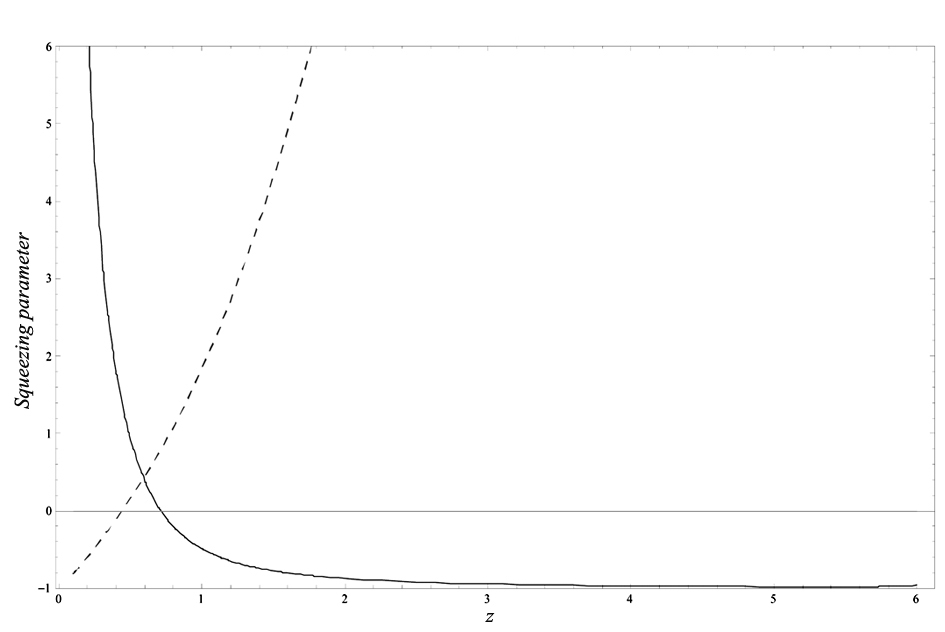}
	\caption{Plots of $S_\phi$ (solid curve), $S_N$ (dashed curve) against $z\in \R$ for three-dimensional harmonic oscillator.}
	\label{fig:figure7}
\end{figure}

\begin{figure}
	\centering
		\includegraphics[width=0.80\textwidth]{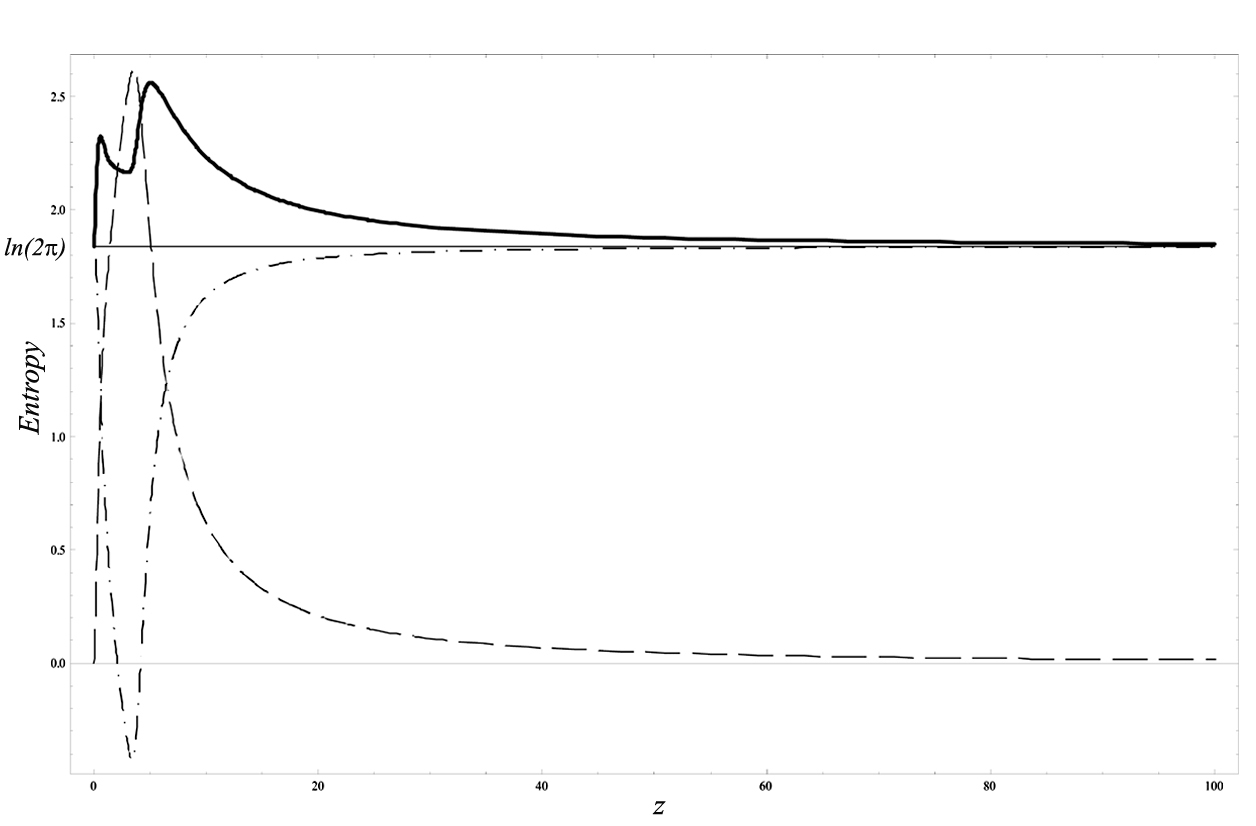}
	\caption{Plots of $R_\phi$ (dot-dashed curve), $R_N$ (dashed curve) and their sum (solid curve) against $z\in \R$ for three-dimensional harmonic oscillator.}
	\label{fig:figure8}
\end{figure}
\begin{figure}
	\centering
		\includegraphics[width=0.70\textwidth]{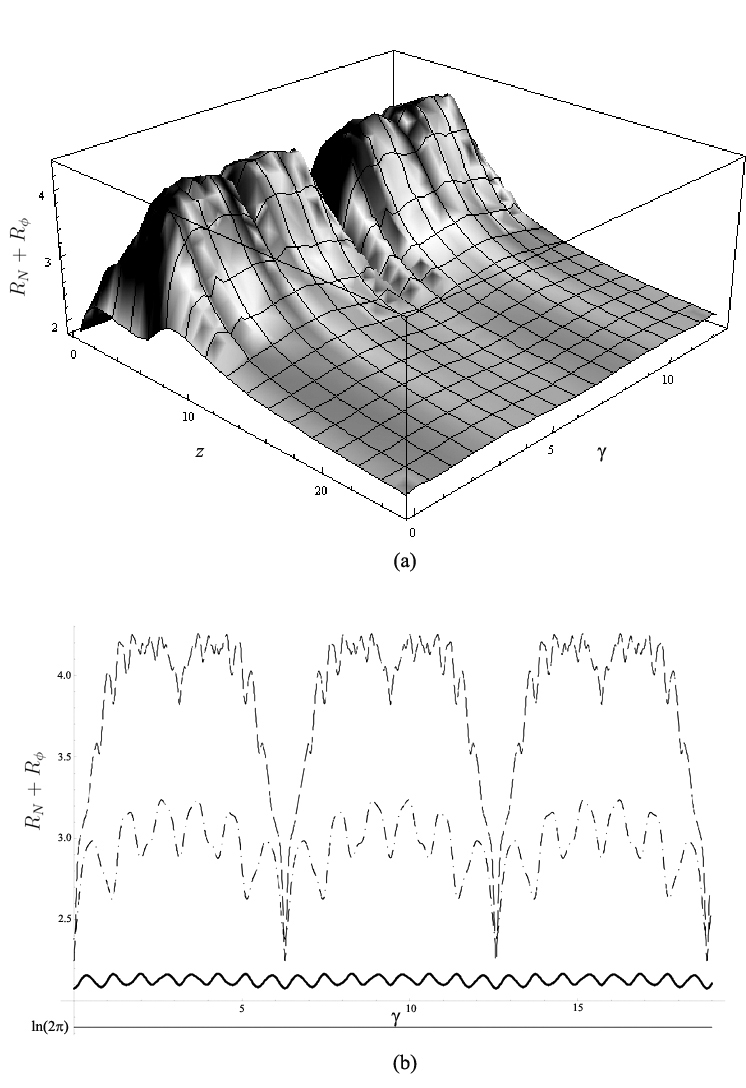}
	\caption{(a) Three-dimensional graph of $R_N+R_\phi$ for a particle in a two-dimensional quantum square box. (b) Plot of $R_N+R_\phi$ against $\gamma$ with $z = 2$ (dot-dashed curve), $z = 5$ (dashed curve) and $z = 20$ (solid curve) for a particle in a two-dimensional quantum square box.}
	\label{fig:figure9}
\end{figure}
\begin{figure}
	\centering
		\includegraphics[width=0.70\textwidth]{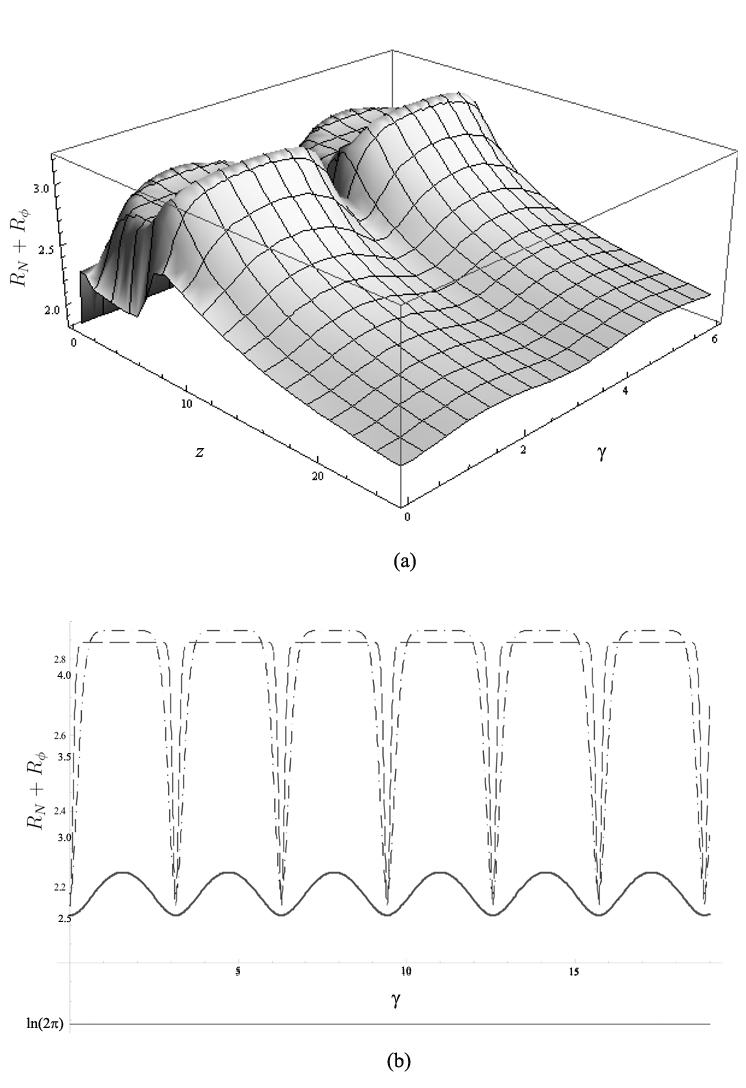}
	\caption{(a) Three-dimensional graph of $R_N+R_\phi$ for three-dimensional harmonic oscillator. (b) Plot of $R_N+R_\phi$ against $\gamma$ with $z = 2$ (dot-dashed curve), $z = 5$ (dashed curve) and $z = 20$ (solid curve) for three-dimensional harmonic oscillator.}
	\label{fig:figure10}
\end{figure}

 
 \end{document}